\documentclass[pre,aps,reprint,twocolumn,amsmath]{revtex4}

\usepackage{graphicx}

\begin{document}

\title{Optimal cooperation-trap strategies for the 
  iterated Rock-Paper-Scissors game \footnote{HJZ
    conceived and performed research; ZB performed numerical simulations and
    checked independently the theoretical results; HJZ wrote the paper.
    Correspondence should be addressed to HJZ ({\tt zhouhj@itp.ac.cn}).}
}

\author{Zedong Bi and Hai-Jun Zhou}
\affiliation{
  State Key Laboratory of Theoretical Physics,
  Institute of Theoretical Physics, Chinese Academy of Sciences, 
  Beijing 100190, China
}

\date{June 16, 2014}

\begin{abstract}
  In an iterated non-cooperative game, if all the players act to maximize 
  their individual accumulated payoff, the system as a whole usually converges
  to a Nash equilibrium that poorly benefits any player.
  Here we show that such an
  undesirable destiny is avoidable in an iterated Rock-Paper-Scissors (RPS)
  game involving two players X and Y. Player X has the option of  proactively
  adopting a cooperation-trap 
  strategy, which enforces complete cooperation from the rational player Y and
  leads to a highly beneficial as well as maximally fair
  situation to both players.
  That maximal degree of cooperation is achievable in such a competitive
  system with cyclic dominance of actions may stimulate creative thinking
  on how to resolve conflicts and enhance cooperation
  in human societies.
\end{abstract}

\maketitle


\section{Introduction}

The solution concept of Nash equilibrium (NE) plays a fundamental role 
both in classic game theory and in evolutionary game theory
\cite{Nash-1950,Osborne-Rubinstein-1994,MaynardSmith-Price-1973,Weibull-1995}.
This concept is developed under the assumption that the players of a
game system are sufficiently rational, so that they are able to 
learn accurately the strategies of the competing players and 
to optimize their own strategy accordingly.
A Nash equilibrium is then a point in the strategy space of the game system
such that any single player is unable to achieve better performance by
changing her/his own strategy in any arbitrary way.

Many non-cooperative games have only a unique NE.
When such a game is played by highly rational players who act to maximize
their individual accumulated payoff, 
it is unavoidable that the system will sooner or later converge to
this unique equilibrium situation.
Unfortunately, however, it is usually the case that the
NE of a non-cooperative game is an unfavorable or even
miserable destiny for all the players. 
Let's consider the two-player Prisoner's Dilemma (PD)
game as a simple example. The cooperative situation of both players choosing not
to confess is much better than the defection situation of both players choosing
to confess, but the latter is the unique NE of this game while
the former is not \cite{Fisher-2008}.
The Nash equilibrium theory therefore predicts that cooperation is unlikely to
sustain when rational players face the conflict between self-interest and group
benefit. Yet cooperation is actually a ubiquitous phenomenon of human society
at all levels,  and it is also widely observed in various biological systems.
Researchers have been puzzled by these facts very much for
many years, and they have proposed a long list of microscopic 
mechanisms trying to explain
the promotion and maintenance of cooperation
\cite{Axelrod-1984,Kollock-1998,Nowak-Highfield-2011}.

In this paper we study cooperation in the itererated two-player
Rock-Paper-Scissors (RPS) game, which is a fundamental non-cooperative
game with cyclic dominance among its action choices (namely
Rock beats Scissors, Scissors beats Paper, and Paper in turn beats
Rock), see Fig.~\ref{fig:ncmodel}.
While the NE theoretical framework assumes that
the rational players of such a game behave \emph{passively} in the
sense that they try to maximize individual gains by
making best responses to the inferred/experienced strategy of the
opponent,  we assume that one of the players might 
act more \emph{proactively}. An intelligent and rational
player may ask the following question: how should I
design my own strategy so that my rational opponent(s), in best response to me, 
for sure will adopt certain strategy that is most beneficial to me?
In later discussions we refer to such a strategy as a cooperation-trap (CT) 
strategy, as it has the
effect of trapping an opponent in a cooperation state. When optimized, such
a CT strategy offers high and maximally fair accumulated payoffs to both
players.

In a literature search for related studies, we found that an early paper
of Grofman and Pool \cite{Grofman-Pool-1977} investigated cooperation in
the PD game from the same angle of intelligent design. 
In this pioneer but largely forgotten
paper, the authors proved that a partial Tit-for-Tat strategy
\cite{Rapoport-Chammah-1965} has the 
potential of enforcing cooperation in the two-player iterated PD game.
The Win-stay, Lose-shift strategy
\cite{Kraines-Kraines-1993,Nowak-Sigmund-1993} can also be analyzed in a
similar way.

The present effort can be regarded as an extension of the 
Grofman-Pool theory to the iterated RPS game, which has the additional 
difficulty of
having more than two action choices that are related by a rotation 
symmetry (see Fig.~\ref{fig:ncmodel}B).
This same theoretical framework may also be applicable to 
many other two-player iterated non-cooperative games, and it may
serve as a guiding principle of designing fair solutions or
strategies for the purpose of
resolving conflicts and enhancing cooperation in human societies.

\section{The Rock-Paper-Scissors game}
 
\begin{figure}[t]
  \begin{center}
    \includegraphics[width=0.3\textwidth]{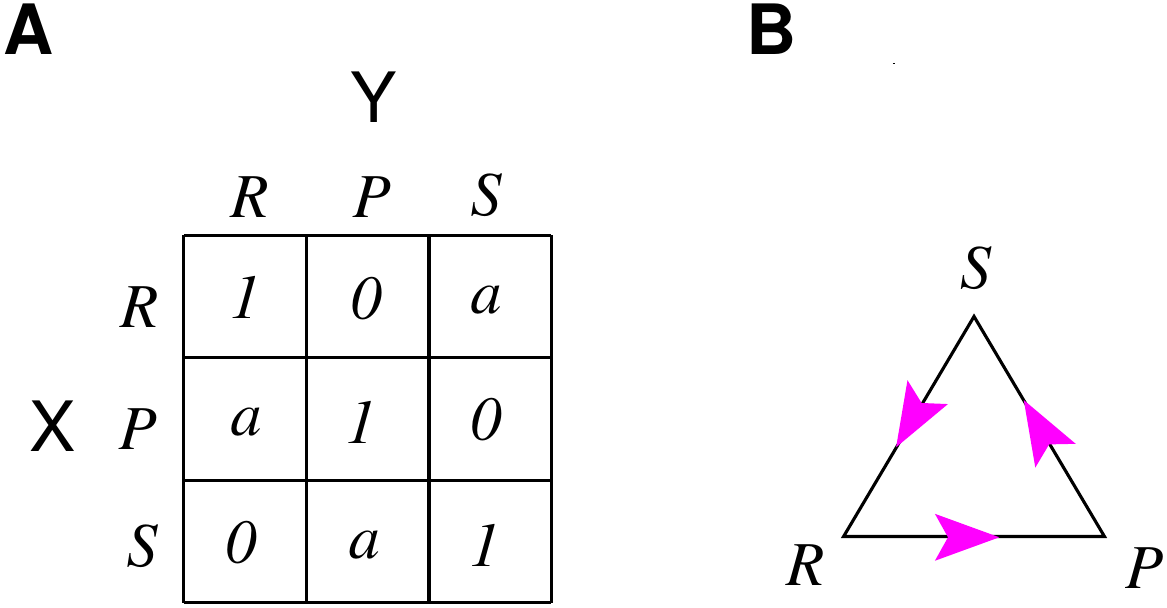}
  \end{center}
  \caption{\label{fig:ncmodel}
    {\bf The RPS game.}
    (A) The payoff matrix. 
    Each matrix element is the payoff of the row player X's action in
    competition with the column player Y's action. (B) 
    The cyclic (non-transitive) dominance relationship among the three
    candidate actions: Rock ($R$) beats Scissors ($S$), $S$ beats Paper ($P$),
    and $P$ in turn beats $R$.
  }
\end{figure}

Consider two players X and Y  playing the RPS game for an
indefinite number of rounds.  At every game round each player can 
choose one action among 
three candidate actions $R$ (rock), $P$ (paper) and $S$
(scissors).
This game has only a single parameter, the payoff $a$ ($>1$) of the
winning action (see Fig.~\ref{fig:ncmodel}A). 
For example, if the 
player X chooses action $R$ in one game round and her
opponent Y chooses action $S$, then X wins with
payoff $a$ and Y loses and gets zero payoff; if the competition is a
tie  with both players choosing the same action, each
player gets unit payoff.

When $a>1$ the system has only a unique NE and it is mixed-strategy in nature,
namely each player chooses the three actions with equal probability $1/3$ 
at every game round independently of each other 
and of the prior action choices \cite{Osborne-Rubinstein-1994}. 
In this mixed-strategy NE the expected payoff per round (EPR) 
for each player is then simply $g_0\equiv (1+a)/3$.
We refer to $g_0$  as the NE payoff.
For $1 < a < 2$ the NE payoff is less than the unit payoff value
each player would get if both players choose the same action in every
game round, and consequently the NE is evolutionarily unstable in this
parameter range. When $a>2$ the NE mixed strategy outperforms the
pure strategy of both players choosing the same action,
and the NE is then evolutionarily stable
\cite{MaynardSmith-Price-1973,Taylor-Jonker-1978,Weibull-1995} and is
the converging point of various dynamical learning 
processes \cite{Sandholm-2010}.

\subsection{Memoryless cooperation-trap strategies}

We now develop CT strategies for player X, and begin with the simplest case of
memoryless strategies, namely at every game round player X does not consider her
and her opponent's prior actions nor the outcomes of prior plays 
but chooses actions $R$, $P$ and $S$ according to the corresponding 
probabilities
$p_r$, $p_p$, and $p_s$ ($\equiv 1- p_r - p_p$), which are fixed by player X
at the beginning of the whole game. Without loss of generality we assume that
$p_r \geq p_p$ and $p_r \geq p_s$, i.e., action $R$ is a favoriate
choice of  X.

As player Y is sufficiently intelligent, he will figure out the strategy of X
after a small number of game repeats. (Alternatively,
player X may also explicitly inform Y about her two choice parameters 
$p_r$ and $p_p$.)
And since Y is sufficiently rational, he then for sure will adopt the
optimized probabilities $q_r^{*}$, $q_p^{*}$, and $q_s^{*}$ 
($\equiv 1-q_r^{*}-q_p^{*}$) of choosing the three actions $R$, $P$ and $S$. The
EPR $g_x$ of player X and the optimized EPR $g_y^{*}$ 
of player Y are 
\begin{eqnarray}
  g_x  =  
  p_r q_r^{*} + p_p q_p^{*} + p_s q_s^{*} 
  + a ( p_r q_s^{*} + p_p q_r^{*} + p_s q_p^{*}) \; , \\
  g_y^{*}  =  
  q_r^{*} p_r + q_p^{*} p_p + q_s^{*} p_s + 
a ( q_r^{*} p_s + q_p^{*} p_r + q_s^{*} p_p ) \; .
\end{eqnarray}

If the strategy of player X have the following
property that $p_r \geq p_s > p_p$,
then because action $P$ is strictly the least favored choice of
player X, then player Y realizes that it is of his best interest
to choose action $R$ in every game round 
($q_r^{*}=1, q_p^{*}=q_s^{*}=0$)  if $p_r - p_p > a (p_r - p_s)$ but
to choose action $P$ in every game round ($q_p^{*} = 1, q_s^{*}=q_r^{*}=0$) if 
$p_r - p_p < a (p_r - p_s)$. In other words, player X traps player Y to
stay in a pure strategy which has maximal degree of predictability. Player
X of course should choose the strategy parameters $p_r$ and $p_p$ to maximize
her EPR $g_x$ under the constraint of not destroying the nice 
trapping effect of her strategy. 
It is not difficult to verify the following conclusions:
(1)  If the payoff parameter $a \in [1, (1+\sqrt{3})/2)$,
  the optimal CT strategy is
  \begin{equation}
    \label{eq:optA}
    p_r^{*} = \frac{a}{2 a-1} -\epsilon \;, \quad
    p_p^{*} = 0 \; , \quad
    p_s^{*} = \frac{a-1}{2 a-1} + \epsilon \; .
  \end{equation}
  (Here and in  latter discussions, 
  $\epsilon \rightarrow 0^{+}$ is an arbitrarily small
  positive value.) The associated maximal EPR  for player X is
  $g_x^{*} = a/(2 a - 1)$, while player Y is very satisfied with sticking to
  action $R$ and getting a larger EPR  of 
  $g_y^{*} = a^2 /(2 a - 1)$. To give a concrete example,
  at $a=1.1$ we have $g_x^{*}\approx 0.917$ and $g_y^{*} \approx 1.008$,
  which are considerably larger than the NE payoff $g_0 = 0.7$.
  (2)    If $a \in (2  + \sqrt{3}, +\infty)$,
  the optimal CT strategy is
  \begin{equation}
    \label{eq:optB}
    p_r^{*}=\frac{a}{2 a-1} + \epsilon\;, \quad 
    p_p^{*}=0\; ,  \quad
    p_s^{*} = \frac{a-1}{2 a - 1} - \epsilon \; ,
  \end{equation}
  The associated
  optimal EPR of player X is $g_x^{*} =  a (a-1) / (2 a - 1)$, while
  player Y receives a larger EPR value of $g_y^{*} =  a^2 / (2 a -1)$
  by sticking to action $P$. Notice that when $a$ is sufficiently
  large, $g_x^{*} \approx a/2-1/4$
  and $g_y^{*} \approx a/2 + 1/4$, which are almost $1.5$ times that of
  the NE payoff $g_0$.

\begin{figure}[t]
  \begin{center}
    \includegraphics[width=0.65\textwidth,angle=270]{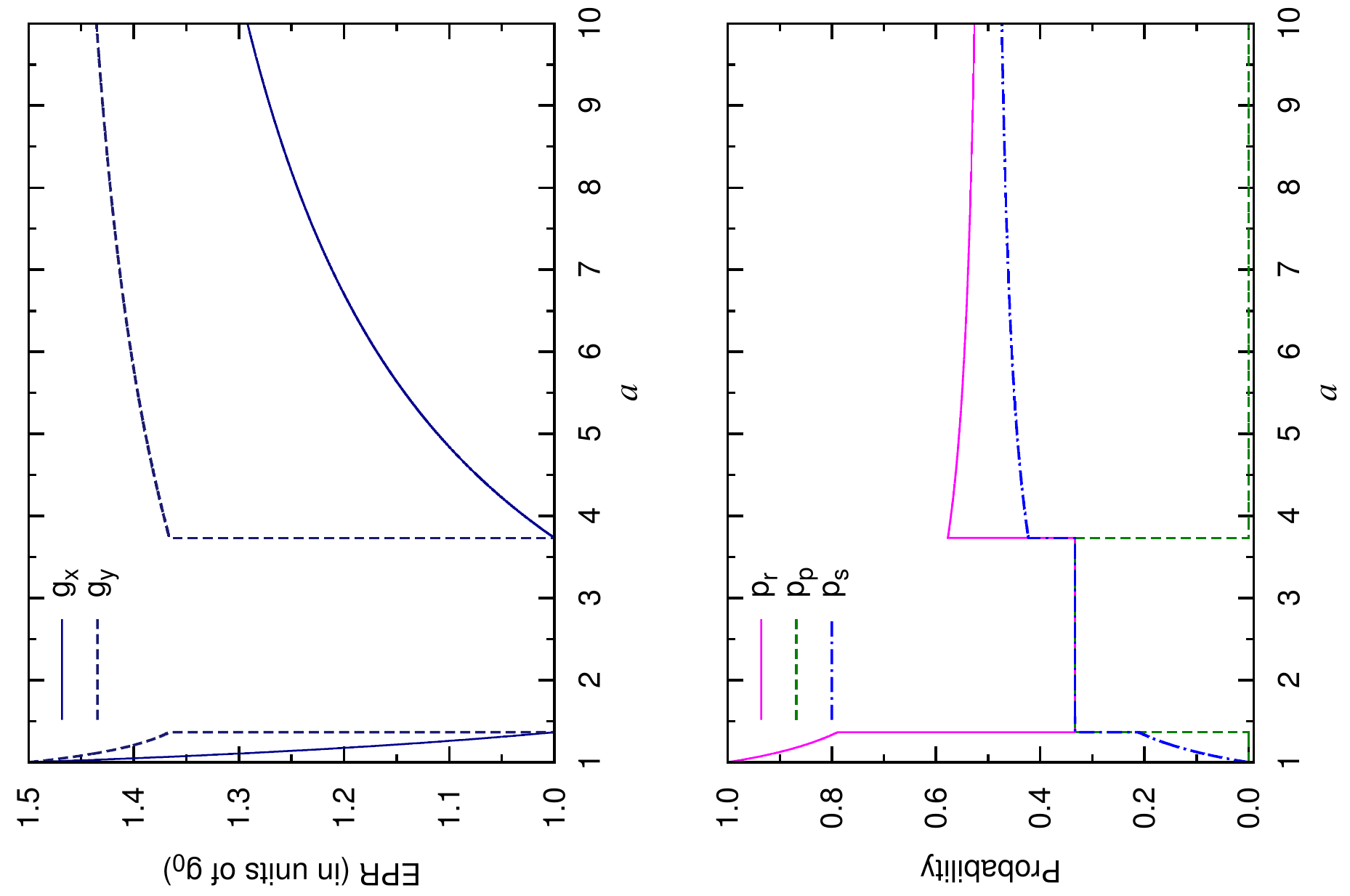}
  \end{center}
  \caption{\label{fig:opt}
    {\bf Optimal memoryless CT strategy.}
    The optimal values of both players' EPR $g_x$ and 
    $g_y$ are shown in the upper panel 
    (in units of NE payoff $g_0$) for each fixed value of $a$, 
    while the optimal values of the
    CT strategy's choice probabilities $p_r$, $p_p$ and $p_s$ are shown
    in the lower panel. When $a\in [1.366, 3.732]$ the NE mixed strategy
    is better for player X than the CT strategy. 
  }
\end{figure}

Figure \ref{fig:opt} gives a direct view about how the optimal
EPRs of both players  and the optimal CT strategy of player X
change with $a$. This optimal memoryless CT strategy indeed offers both
players higher accumulated payoffs than the NE mixed strategy does.
However, the passive player Y benefits more than the proactive player X.
It is then natural for player X to feel that she has sacrificed
too much for enforcing cooperation and to declare that such a CT strategy,
although better than the NE mixed strategy, is unfair as her
opponent earns more by free riding. Furthermore, this CT
strategy is worse than the NE mixed strategy in the parameter range of
$a \in [1.366, 3.732]$.

These shortcomings of the memoryless CT strategy
can be eliminated by increasing
the memory length of the CT strategy.

\subsection{Cooperation-trap strategies with finite memory length}

Recent laboratory experiments of Wang and Xu \cite{Wang-Xu-Zhou-2014} revealed
that decision-making of human subjects has strong memory effect, namely the
payoffs of the previous game rounds influence considerably
a player's action choices in the following game rounds. 
For the RPS game, the implications of such conditional response strategies
have not yet been fully explored. Here we suggest that the proactive player
X can adopt an optimized version of such a strategy to enforce fair cooperation.

\begin{figure}[t]
  \begin{center}
    \includegraphics[width=0.65\textwidth,angle=270]{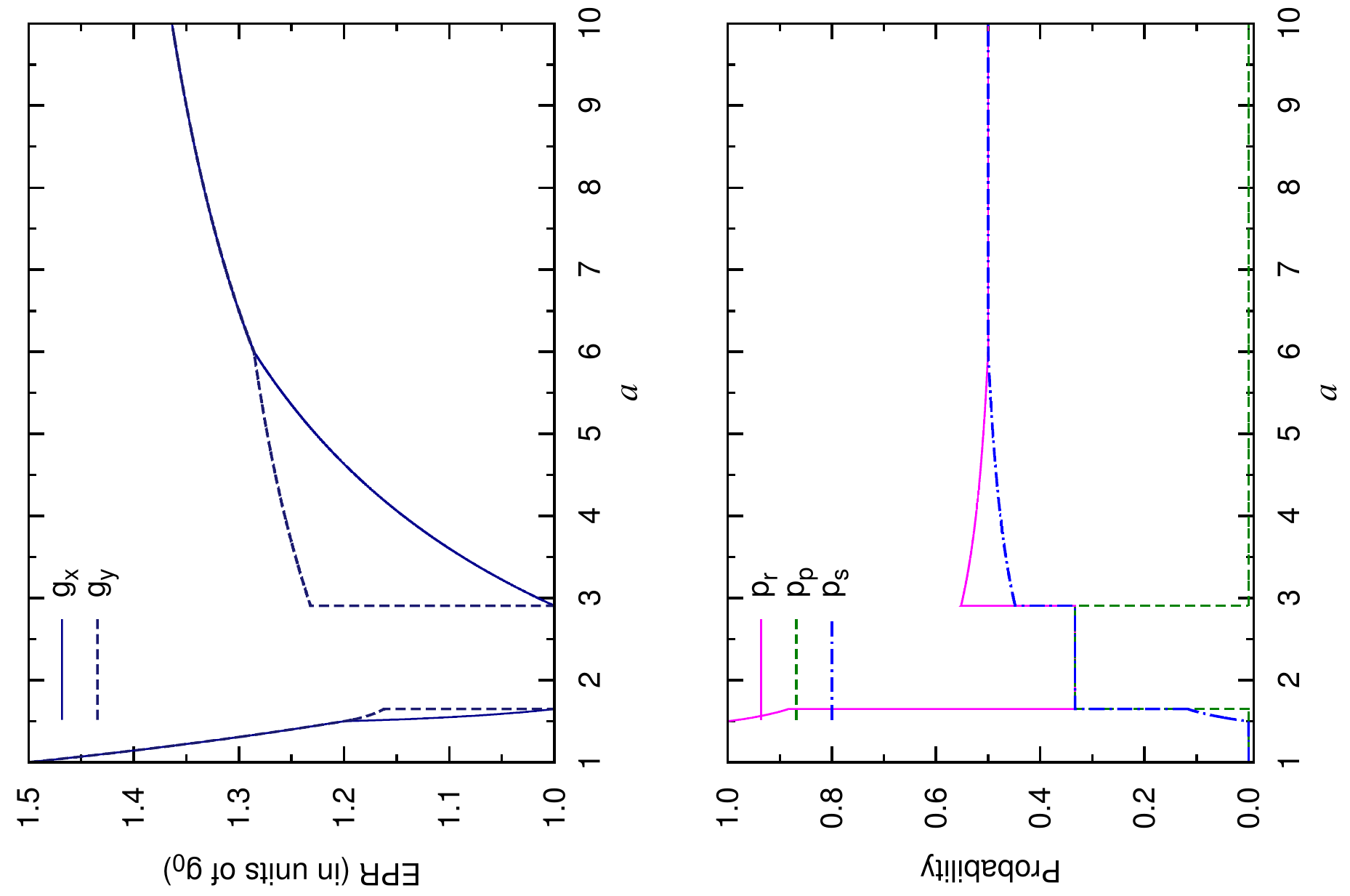}
  \end{center}
  \caption{\label{fig:CRopt}
    {\bf Optimal CT strategy of unit memory length.}
    The optimal values of both players' EPR $g_x$ and 
    $g_y$ are shown in the upper panel 
    (in units of NE payoff $g_0$) for each fixed value of $a$, 
    while the optimal values of the
    CT strategy's choice probabilities $p_r$, $p_p$ and $p_s$ are shown
    in the lower panel. When $a\in [1.649, 2.905]$ the NE mixed strategy
    is better for player X than the CT strategy.
  }
\end{figure}

When the payoff parameter $a>2$, a play output of win-lose brings
payoff $a$ to
the group, while a tie output only brings lower payoff $2$. Therefore it is
desirable for player X to discourage the occurrence of tie output. For the 
simplest case of unit memory length, the CT strategy then goes as follows:
If player X wins over or loses to player Y in the previous game round,
then in the next round she chooses action $R$ with probability $p_r \geq 1/2$ 
and action $S$ with probability $p_s=1-p_r$ (she avoid choosing action $P$, 
i.e., $p_p=0$); but if X ties with Y in the previous game round, 
then in the next round she chooses the three candidate actions with equal
probability $1/3$. This strategy has only a single parameter $p_r$. 
The motivation for player X to adopt the NE mixed strategy after experiencing
a tie output is to discourage player Y from choosing action $R$:
although Y might get a higher expected payoff in one game round by 
choosing action
$R$ rather than action $P$, the former choice has a high probability 
of leading to
a tie, which will then reduce player Y's expected payoff to $g_0$
in the following one or even more game rounds.

On the other hand, when $1<a<2$, a play output of tie is better off to the
group than a win-lose output. Then player X  has the option of implementing
a CT strategy to discourage player Y from either
winning over or losing to her. Again for the simplest case of unit memory
length, the recipe of the CT strategy is:
If player X ties with player Y in the previous game round,
then in the next round she chooses action $R$ with probability $p_r \geq 1/2$ 
and action $S$ with probability $p_s=1-p_r$;
but if X either wins over or loses to Y in the previous round, 
then in the next round she chooses the three candidate actions with
equal probability $1/3$.

It turns out that the optimal CT strategy of
unit memory length has the following quantitative properties:
(1) If $ a \in [1,  3/2]$, then the optimal value $p_r^{*}$ for the
choice probability $p_r$ is $p_r^{*}=1-\epsilon$,
and the optimal EPRs of player X and player Y are equal, 
$g_x^{*}=g_y^{*}=1$.
(2) If $a \in (3/2,  1.649)$, then 
$p_r^{*} = 8 a /[9 a -3 + \sqrt{(9 a - 3)^2 - 48 a^2}]-\epsilon$,
and the optimal EPRs for X and Y are, respectively,
$g_x^{*}=[1+a (1-p_r^{*})]/(4- 3 p_r^{*})$ and 
$g_y^{*}=[1+ 2 a (1-p_r^{*})]/(4-3 p_r^{*})$.
(3) If $a \in (2.905,    6)$, then 
$p_r^{*} = 4 a/[ 3 (a-1)+\sqrt{9 (a-1)^2 +24 a^2}] + \epsilon$, 
and the optimal EPRs for X and Y are, respectively,
$g_x^{*}=(1-p_r^{*}) a$ and $g_y^{*} = p_r^{*} a$.
(4) If $a \in [6, +\infty)$, then  $p_r^{*}=1/2 + \epsilon$,
  and the optimal EPRs
  for X and Y are eqaul,  $g_x^{*}=g_y^{*}= a/2$.
Figure~\ref{fig:CRopt} gives a direct view of these properties.
Compared with the optimal memoryless CT strategy of Fig.~\ref{fig:opt}, 
we notice a major qualitative
improvement is that this new optimal CT strategy
achieves fair outcomes to player X and player Y
when $1\leq a \leq 3/2$ or $a\geq 6$. However, this optimal CT strategy of
unit memory length is still not perfect, as it is not applicable for
$a \in [1.649, 2.905]$, and
it is not completely fair to the proactive player X for $ a \in (3/2, 1.649)
\cup (2.905, 6)$.

\begin{figure}[t]
  \begin{center}
    \includegraphics[width=0.65\textwidth,angle=270]{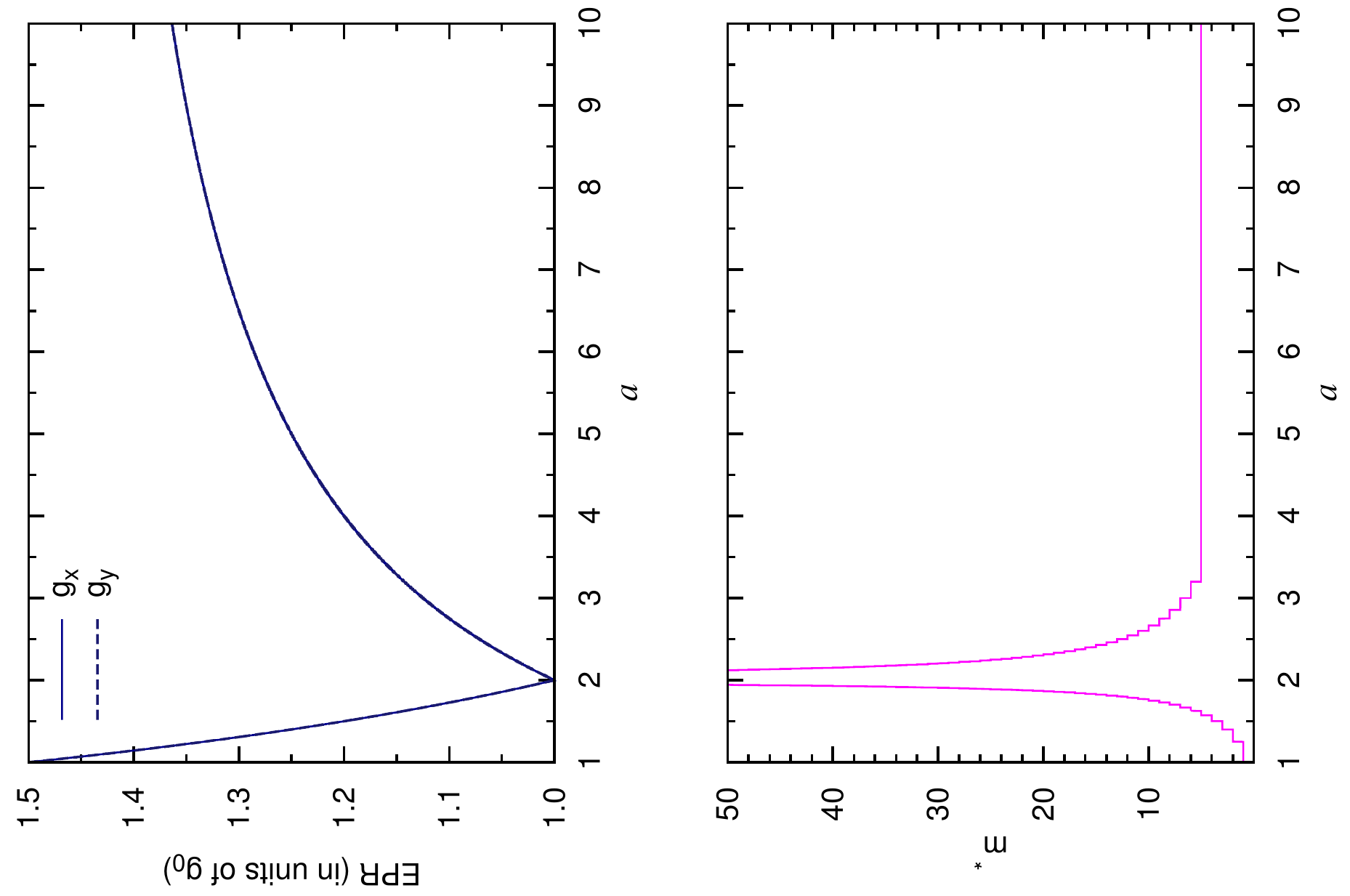}
  \end{center}
  \caption{\label{fig:CRoptM}
    {\bf Optimal CT strategy of finite memory length.}
    The optimal values of both players' EPR $g_x$ and 
    $g_y$ are shown in the upper panel 
    (in units of NE payoff $g_0$) for each fixed value of $a$, 
    while the minimal memory length $m^{*}$ of the CT strategy
    is shown in the lower panel.
  }
\end{figure}

To completely eliminate these undesirable features, player
X can increase the memory length of her CT strategy and therefore be more 
non-tolerant to defection. There are many ways of implementing
such an idea. When $a>2$, arguably the simplest CT strategy of memory
length $m$ 
goes as follows: By default player X adopts the mixed 
strategy $(p_r, 0, 1-p_r)$ in 
every game round, namely she chooses action $R$ with probability 
$p_r \geq 1/2$ and
action $S$ with the remaining
probability $p_s=1-p_r$; however if a tie occurs in one game
round, then player X shifts to the NE mixed strategy $(1/3, 1/3, 1/3)$ in 
the next $m$ game rounds and then shifts back to the default strategy
$(p_r, 0, 1-p_r)$ in the $(m+1)$-th game round. 
It is a simple exercise to check that, if
\begin{equation}
\label{eq:crmopt}
p_r \geq \max\Bigl(\frac{1}{2} ,
\frac{2 a}{\sqrt{4 m a^2 + A^2}-|A|}
\Bigr) \; ,
\end{equation}
where $A\equiv 1+(1+a) m /3 - 2 a$,
then player Y will be satisfied with sticking to action $P$ in every game round.
If player X sets the memory length to
the smallest positive integer $m^{*}$ which reduces Eq.~(\ref{eq:crmopt})
to the trivial requirement of $p_r \geq 1/2$,
then it is optimal for player X to set $p_r$ to the value
$p_r^{*}=1/2$, and the optimal EPRs  for player X and player Y
are equal, 
$g_x^{*} = g_y^{*} = a/2$. Notice that for $a$ approaches $2$ from above with
$a= 2 + \epsilon$, the required
minimal memory length diverges as $m^{*}\approx 6 / \epsilon$. In other words,
it is most difficult to enforce fair cooperation when $a\approx 2$, see
Fig.~\ref{fig:CRoptM}.

If the payoff parameter $a<2$, an optimal CT strategy with memory length $m$ 
can be constructed following the same line of reasoning as above, namely that
player X adopts action $R$ at every game round, but if she loses to player
Y in one game round, then she shifts to the
NE mixed strategy in the next $m$ game rounds and then shifts back to
the default strategy $(1, 0, 0)$ in the $(m+1)$-th game round.
We can
easily verify that
if player X sets the memory length to be $m > 3(a-1)/(2-a)$, then
it is optimal for player Y to stick to action $R$ in every game round, and
the optimal EPRs for both players are equal, 
$g_x^{*}= g_y^{*}=1$.

As clearly demonstrated in Fig.~\ref{fig:CRoptM}, for each payoff parameter
$a\neq 2$, an optimal CT strategy with a finite memory length $m^*$ can
be implemented to achieve maximal and fair accumulated payoff for both
players. At $a=2$, there is no need to adopt a CT strategy, as the NE
mixed strategy is itself optimal.

\section{Discussion}

We have demonstrated in this paper that fair cooperation
can be achieved in the two-player iterated RPS game.
Such a highly cooperative state brings maximal accumulated payoff
to the group, and it 
is not enforced by external authorities but by the proactive decision of one
player to adopt an optimal cooperation-trap strategy. 
The basic designing principle
of such optimal CT strategies should be generally applicable to
other two-player iterated non-cooperation games. 

For the optimal CT strategies to work, the passive player  Y is assumed to be
considerably rational so that he adopts a best response strategy to that of
his opponent X to maximize his accumulated payoff,
while the proactive player X is assumed  in addition to be wise enough so that
she does not exploit the cooperation state of her opponent too much but is 
satisfied with a fair share of the total accumulated group payoff.
This latter assumption might be a little bit too strong, but maybe it is
not strictly necessary as player Y will punish X for defection behaviors.

For the iterated RPS game, it appears to be
impossible for the proactive player X to
design a CT strategy which brings  higher expected payoff per game
round to herself than to her opponent. 
However, this is not a general conclusion. For some other
game systems, notably the  iterated PD game 
\cite{Grofman-Pool-1977,Press-Dyson-2012}), the proactive
player X has the option of optimizing her CT strategy
to extort her opponent Y. We do not recommend the adoption of such
greedy strategies, as the opponent player Y  will very likely be frustrated
by the defection behaviors of player X and he may then choose not
to cooperate even such a choice hurts also himself 
\cite{Hilbe-Rohl-Milinski-2014}.

When strategic interactions  occur in biological systems, the involved
individual animals, insects, bacteria, cells, ..., are of course
far from being rational or sufficiently intelligent. However the collective
decision-making of such agents at the population level, aided by the 
evolutionary mechanism of mutation and selection, may appear to be 
very rational. By trial
and error, such systems may develop certain CT-like strategies even without
the need of intelligent designing. 
It would be very interesting to investigate empirically
whether CT strategies are actually implemented in some biological
systems.

Cooperation in a finite-population RPS game system with more than two
players may be much more difficult to achieve than the case of
two players. A recent theoretical investigation by one of the
present authors \cite{Wang-Xu-Zhou-2014} suggested that optimized
conditional response strategies might offer higher accumulated payoffs
to individual players than the NE mixed strategy does. But it is still
an open question as to whether high degree of coopeation can also be
enforced in a multiple-player iterated RPS game by a number of
proactive players. 
We leave such a challenging issue to future investigations.

$\;$
\vskip 2.0cm

\section*{Acknowledgments}

HJZ was supported by the National Basic Research Program
of China (2013CB932804)
and the National Science Foundation of China
(11121403, 11225526).
HJZ thanks Zhijian Wang and Bin Xu for a recent fruitful
collaboration on the finite-population Rock-Paper-Scissors game, which
inspired the present work greatly.


\begin{thebibliography}{10}
\expandafter\ifx\csname url\endcsname\relax
  \def\url#1{\texttt{#1}}\fi
\expandafter\ifx\csname urlprefix\endcsname\relax\def\urlprefix{URL }\fi
\providecommand{\bibinfo}[2]{#2}
\providecommand{\eprint}[2][]{\url{#2}}

\bibitem{Nash-1950}
\bibinfo{author}{Nash, J.~F.}
\newblock \bibinfo{title}{Equilibrium points in $n$-person games}.
\newblock \emph{\bibinfo{journal}{Proc. Natl. Acad. Sci. USA}}
  \textbf{\bibinfo{volume}{36}}, \bibinfo{pages}{48--49}
  (\bibinfo{year}{1950}).

\bibitem{Osborne-Rubinstein-1994}
\bibinfo{author}{Osborne, M.~J.} \& \bibinfo{author}{Rubinstein, A.}
\newblock \emph{\bibinfo{title}{A Course in Game Theory}}
  (\bibinfo{publisher}{MIT Press}, \bibinfo{address}{New York},
  \bibinfo{year}{1994}).

\bibitem{MaynardSmith-Price-1973}
\bibinfo{author}{{Maynard Smith}, J.} \& \bibinfo{author}{Price, G.~R.}
\newblock \bibinfo{title}{The logic of animal conflict}.
\newblock \emph{\bibinfo{journal}{Nature}} \textbf{\bibinfo{volume}{246}},
  \bibinfo{pages}{15--18} (\bibinfo{year}{1973}).

\bibitem{Weibull-1995}
\bibinfo{author}{Weibull, J.~M.}
\newblock \emph{\bibinfo{title}{Evolutionary Game Theory}}
  (\bibinfo{publisher}{MIT Press}, \bibinfo{address}{Cambridge, MA},
  \bibinfo{year}{1995}).

\bibitem{Fisher-2008}
\bibinfo{author}{Fisher, L.}
\newblock \emph{\bibinfo{title}{Rock, Paper, Scissors: Game Theory in Everyday
  Life}} (\bibinfo{publisher}{Basic Books}, \bibinfo{address}{New York},
  \bibinfo{year}{2008}).

\bibitem{Axelrod-1984}
\bibinfo{author}{Axelrod, R.}
\newblock \emph{\bibinfo{title}{The Evolution of Cooperation}}
  (\bibinfo{publisher}{Basic Books}, \bibinfo{address}{New York},
  \bibinfo{year}{1984}).

\bibitem{Kollock-1998}
\bibinfo{author}{Kollock, P.}
\newblock \bibinfo{title}{Social dilemmas: The anatomy of cooperation}.
\newblock \emph{\bibinfo{journal}{Annu. Rev. Sociol.}}
  \textbf{\bibinfo{volume}{24}}, \bibinfo{pages}{183--214}
  (\bibinfo{year}{1998}).

\bibitem{Nowak-Highfield-2011}
\bibinfo{author}{Nowak, N.~A.} \& \bibinfo{author}{Highfield, R.}
\newblock \emph{\bibinfo{title}{SuperCooperators: Altruism, Evolution, and Why
  We Need Each Other to Succeed}} (\bibinfo{publisher}{Free Press},
  \bibinfo{address}{New York}, \bibinfo{year}{2011}).

\bibitem{Grofman-Pool-1977}
\bibinfo{author}{Grofman, B.} \& \bibinfo{author}{Pool, J.}
\newblock \bibinfo{title}{How to make cooperation the optimizing strategy in a
  two-person game}.
\newblock \emph{\bibinfo{journal}{J. Math. Sociol.}}
  \textbf{\bibinfo{volume}{5}}, \bibinfo{pages}{173--186}
  (\bibinfo{year}{1977}).

\bibitem{Rapoport-Chammah-1965}
\bibinfo{author}{Rapoport, A.} \& \bibinfo{author}{Chammah, A.~M.}
\newblock \emph{\bibinfo{title}{Prisoner's Dilemma: A Study in Conflict and
  Cooperation}} (\bibinfo{publisher}{University of Michigan Press},
  \bibinfo{address}{Ann Arbor, Michigan}, \bibinfo{year}{1965}).

\bibitem{Kraines-Kraines-1993}
\bibinfo{author}{Kraines, D.} \& \bibinfo{author}{Kraines, V.}
\newblock \bibinfo{title}{Learning to cooperate with pavlov: an adaptive
  strategy for the iterated prisoner's dilemma with noise}.
\newblock \emph{\bibinfo{journal}{Theory and Decision}}
  \textbf{\bibinfo{volume}{35}}, \bibinfo{pages}{107--150}
  (\bibinfo{year}{1993}).

\bibitem{Nowak-Sigmund-1993}
\bibinfo{author}{Nowak, M.} \& \bibinfo{author}{Sigmund, K.}
\newblock \bibinfo{title}{A strategy of win-stay, lose-shift that outperforms
  tit-for-tat in the prisoner's dilemma game}.
\newblock \emph{\bibinfo{journal}{Nature}} \textbf{\bibinfo{volume}{364}},
  \bibinfo{pages}{56--58} (\bibinfo{year}{1993}).

\bibitem{Taylor-Jonker-1978}
\bibinfo{author}{Taylor, P.~D.} \& \bibinfo{author}{Jonker, L.~B.}
\newblock \bibinfo{title}{Evolutionarily stable strategies and game dynamics}.
\newblock \emph{\bibinfo{journal}{Mathematical Biosciences}}
  \textbf{\bibinfo{volume}{40}}, \bibinfo{pages}{145--156}
  (\bibinfo{year}{1978}).

\bibitem{Sandholm-2010}
\bibinfo{author}{Sandholm, W.~M.}
\newblock \emph{\bibinfo{title}{Population Games and Evolutionary Dynamics}}
  (\bibinfo{publisher}{MIT Press}, \bibinfo{address}{New York},
  \bibinfo{year}{2010}).

\bibitem{Wang-Xu-Zhou-2014}
\bibinfo{author}{Wang, Z.}, \bibinfo{author}{Xu, B.} \& \bibinfo{author}{Zhou,
  H.-J.}
\newblock \bibinfo{title}{Social cycling and conditional responses in the
  rock-paper-scissors game}.
\newblock \bibinfo{howpublished}{arXiv:1404.5199} (\bibinfo{year}{2014}).

\bibitem{Press-Dyson-2012}
\bibinfo{author}{Press, W.~H.} \& \bibinfo{author}{Dyson, F.~J.}
\newblock \bibinfo{title}{Iterated prisoner's dilemma contains strategies that
  dominate any evolutionary opponent}.
\newblock \emph{\bibinfo{journal}{Proc. Natl. Acad. Sci. USA}}
  \textbf{\bibinfo{volume}{109}}, \bibinfo{pages}{10409--10413}
  (\bibinfo{year}{2012}).

\bibitem{Hilbe-Rohl-Milinski-2014}
\bibinfo{author}{Hilbe, C.}, \bibinfo{author}{R\"{o}hl, T.} \&
  \bibinfo{author}{Milinski, M.}
\newblock \bibinfo{title}{Extortion subdues human players but is finally
  punished in the prisoner's dilemma}.
\newblock \emph{\bibinfo{journal}{Nature Commun.}}
  \textbf{\bibinfo{volume}{5}}, \bibinfo{pages}{3976} (\bibinfo{year}{2014}).

\end{thebibliography}

\end{document}